\documentstyle[prd,aps,preprint,tighten,floats]{revtex}

\newcommand{\rf}[1]{(\ref{eq:#1})}
\newcommand{\R}{{\cal R}}
\newcommand{\Y}{{\cal Y}}
\newcommand{\N}{{\cal N}}
\newcommand{\G}{{\cal G}}

\begin{document}
\author{ M. D. Maia\thanks{E-Mail: maia@mat.unb.br}\\
Universidade de Bras\'{\i}lia, Departamento de Matem\'{a}tica\\
Bras\'{\i}lia, DF. 70910-900}
\title{ Hypersurfaces of Five Dimensional\\ Vacuum space-times}
\maketitle
\draft
\begin{abstract}
It is shown that  a  space-time hypersurface  of a 5-dimensional Ricci flat
space-time  has its  energy-momentum tensor algebraically related
to its extrinsic curvature  and to the Riemann curvature of the embedding
space.  It is  also seen that  the  Einstein-Maxwell field  
does not arise naturally  from this geometry, so that a  Kaluza-Klein model
based on it  would require further assumptions.
\end{abstract}

\vspace{1cm}
It is  well known  that any d-dimensional Riemannian manifold $V_d$ can
be  locally and isometrically embedded into another manifold with a
sufficiently high number of dimensions and  with some specified geometry. The
extension of many  embedding  theorems to 
four-dimensional space-times (regarded as pseudo Riemannian manifolds) is
straightforward. On the other hand, in spite 
of the   many attempts made in the past,  the physical significance 
of  the   extra  degrees of freedom resulting from that embedding, remains
unclear \cite{Fronsdal,Joseph,Ne'emman,Friedman}. 

An obvious physical application of the  embedding of a space-time seems
to be along the lines of Kaluza-Klein theory. 
One,   difficulty of that
theory is its inability to  generate  light chiral fermions in its low energy
sector, so that it fails to pass the consistency test at the 
electroweak level \cite{Niew}. In embedding versions of Kaluza-Klein
and  higher dimensional theories, conjectured by  several authors
\cite{RT,Maia:1,Visser}, 
the extra  dimensions are not necessarily compact. The most likely
candidate to ground state of one such  theory is not necessarily 
like  $M_{4}\times B_{n}$, but rather  a flat embedding space \cite{Maia:1}.
As in standard Kaluza-Klein theory, the higher dimensional space is
a  vacuum solution  (Ricci-flat) of   Einstein's equations, containing the
four dimensional space-time as a  subspace. 
However,  it  would  be necessary to  find another explanation for the non
observability of the extra  dimensions, as for  example by breaking the
translational symmetry  along the  extra dimension\cite{RS}.

In a recent paper \cite{Romero},  a  five dimensional  model along
these lines was studied from the embedding point of view, using
a theorem due to Campbell, according to which  any  space-time of  general
relativity can be  embedded 
in a five dimensional   Ricci-flat space \cite{Campbell}. That result certainly
gives support to the above discussion on the embedding model of Kaluza-Klein
theory.   Campbell's theorem brings also some interesting questions. One
of them concerns the  physical meaning  of the  curvature of the
five dimensional  space. Another  question is, can we obtain a
5-dimensional Kaluza-Klein  unified model?

In this note  we show that 
the energy momentum tensor of  the source of gravitation in four dimensions
is  algebraically  (that is, not  depending on  derivatives) related to  the
extrinsic curvature  of the embedded space-time,  but depends  also on the
Riemann curvature of the 
On the other hand,   it is well known that five dimensional Kaluza-Klein theory
reproduces  Einstein-Maxwell gravitational equations in four dimensions. 
This does not seem to occur in the  five-dimensional Ricci-flat embedding 
program, 
meaning that that if  we want  to built  a  five dimensional embedding model of
unified interactions along the Kaluza-Klein ideas, we may 
need to give up Campbell's result in favor of  a larger number of dimensions.
 
Consider a space-time $V_{4}$ with metric $g_{ij},$ solution of Einstein's
equations for  a given  source
\[
G_{ij}  =R_{ij}-\frac{1}{2}Rg_{ij} = 8\pi G\, t_{ij}^{source},
\]
and its local isometric embedding in a 5-dimensional
manifold ${\cal M}_{5}$  given by the map
$$
{\cal Y} :V_{4}\rightarrow {\cal M}_{5} .
$$
This means that there is  a coordinate chart for ${\cal M}_{5}$ given by 
embedding functions ${\cal Y}^{\mu}(x^{i})$.  If  ${\cal G}_{\mu\nu}$ denotes
the metric of ${\cal M}_{5}$ in the embedding coordinates ${\cal Y}^{\mu}$,
then  together with  a vector 
fields ${\cal N}^{\mu}$, orthogonal to the  space-time, the embedding is 
given by \footnote{Lower case Latin indices run from 1 to 4.  All Greek
indices run from 1 to 5. The 
indicated antissymetrization applies only to the indices of the same kind within
the brackets.} 
\begin{equation}
\label{eq:IMB}g_{ij} ={\cal Y}^{\mu}{}_{,i}{\cal Y}^{\nu}{}_{,j}{\cal G}
_{\mu\nu},\;\;{\cal N}^{\mu}{\cal Y}^{\nu}{}_{,i}{\cal G}
_{\mu\nu}=0,\;\; {\cal N}^{\mu}{\cal N}^{\nu} {\cal G}
_{\mu\nu}=g_{55} =\epsilon =\pm 1 ,
\end{equation}
Notice that the fifth dimension may have space or time-like characters. 
The embedding coordinates $\Y^\mu $ are obtained by integration of  the
above equations.
The components of the second fundamental form (or,
equivalently, the  extrinsic curvature) of  the space-time  may be  written
explicitly as 
\begin{equation}
b_{ij}=- {\cal Y}^\mu {}_{,i}{\cal N}^\nu {}_{;j}{\cal G}_{\mu
\nu }= {\cal Y}^\mu {}_{;ij}{\cal N}^{\nu} {\cal G}_{\mu
\nu }\label{eq:b} 
\end{equation}
This quantity, plus the metric of  $V_{4}$ determines completely the
embedding. Obviously, if the embedding coordinates are
known (as  for example in
\cite{Rosen}), rather than calculated from \rf{IMB}  then all we
have to do is to calculate 
$\N^{\mu}{}$ and $b_{ij}$ \cite{Roque}. However, in doing so 
we learn very little over  what we already know from the intrinsic geometry of
the space-time.
In another, perhaps more meaningful  approach,  the  embedding  conditions
should be  considered as part of the dynamical process. In this case, the
Ricci tensor of $V_{4}$  may be calculated
(as for example in\cite{Collinson,Matsumoto}) directly from the integrability
conditions for the embedding, which are the Gauss and Codazzi  equations:  
\begin{eqnarray}  
R_{ijkl}&=&2 \epsilon b_{i[k}b_{l]j} + {\cal R}_{\alpha\beta\gamma\delta} 
{\cal Z}^{\alpha}_{,i} {\cal Z}^{\beta}_{,j} {\cal Z}^{\gamma}_{,k} {\cal Z}^{\delta}_{,l}
\label{eq:G}.\vspace{3mm}\\
2b_{i[j;k]} &=& {\cal R}_{\alpha\beta\gamma\delta}
{\cal Z}^{\alpha}_{,i} {\cal Z}^{\gamma}_{,j} {\cal Z}^{\delta}_{,k} {\cal N}^{\beta}.
\label{eq:C}
\end{eqnarray}
where ${\cal  R}_{\alpha\beta\gamma\delta}$ is the Riemann tensor of 
$V_{5}$ and  $R_{ijkl}$ denotes that of $V_{4}$. Contraction of \rf{G} with
$g^{il}$  gives 
\[
R_{jk}=\epsilon g^{il}R_{ijkl}= g^{il}(b_{ik}b_{lj}- b_{il}b_{kj}) -
g^{il}{\cal
R}_{\alpha\beta\gamma\delta}\Y^{\alpha}_{,i}\Y^{\beta}_{,j}\Y^{\gamma}_{,k}
\Y^{\delta}_{,l}  
\]
From  (1),  we obtain 
\[
g^{ij}\Y^{\alpha}_{,i}\Y^{\beta}_{,j}= {\G}^{\alpha\beta} -\epsilon {\cal
N}^{\alpha} {\cal N}^{\beta}
\]
Consequently, the  Ricci curvature of the space-time  becomes:
\[
R_{jk}= \epsilon g^{il}(b_{ik}b_{lj}- b_{il}b_{kj})+
{\cal R}_{\beta\gamma}\Y^{\beta}_{,j}\Y^{\gamma}_{,k}
-\epsilon {\cal R}_{\alpha\beta\gamma\delta}
{\cal N}^{\alpha}{\cal N}^{\delta}{\cal Y}^{\beta}_{,j}{\cal Y}^{\gamma}_{,k} 
\]
and the Ricci scalar is
\[
R = \epsilon( k^{2} -h^{2})+{\cal R}-2\epsilon{\cal R}_{\alpha\beta}{\cal
N}^{\alpha}{\cal N}^{\beta} 
\]
where  $k^{2}=b^{mn}b_{mn}$ is the scalar  extrinsic curvature  and
$h=g^{mn}b_{mn}$ is the  mean curvature of  space-time. 
Therefore, the Einstein tensor  of  space-time may be written as
\begin{equation}
G_{jk}=\epsilon t_{jk}^{geom.}  
+ \R_{\beta\gamma}(\Y^{\beta}_{,j}\Y^{\gamma}_{,k}+\epsilon
\N^{\beta}\N^{\gamma} g_{jk})
-\epsilon\, \R_{\alpha\beta\gamma\delta}\N^{\alpha}\N^{\delta}\Y^{\beta}_{,j} 
\Y^{\gamma}_{,k} -\frac{1}{2} {\cal R}g_{jk}  \label{eq:EEE}
\end{equation}
where  we have denoted 
\begin{equation}
t^{geom.}_{jk}=g^{mn}b_{jm}b_{kn} -hb_{jk} -\frac{1}{2} (k^{2}-h^{2})g_{jk}
\label{eq:t}
\end{equation}
Since  the  space-time $V_{4}$  is  a solution of Einstein's equations, 
equations \rf{EEE} tells  that  a  combination of curvatures terms contribute to 
the   energy -momentum tensor of the  gravitational source. 

Now,  we apply the Ricci flatness   condition  for  $V_{5}$,
$\R_{\beta\gamma}=0$ and  $\R=0$. Then from \rf{EEE} and
Einstein's equations  we obtain  
\begin{equation}
G_{ij} =  8\pi G \,t_{ij}^{source}= \epsilon\, t_{ij}^{geom.}-
\epsilon\R_{\alpha\beta\gamma\delta}\N^{\alpha}\N^{\delta}\Y^{\beta}_{,i}  
\Y^{\gamma}_{,j}    \label{eq:EE1}
\end{equation}
These  are the gravitational  equations for  a  space-time
embedded in a Ricci-flat  $V_{5}$. Notice that it does not depend
on derivatives of $b_{ij}$. By  solving  this  equation  algebraically on
$b_{ij}$ together with the constraints  \rf{G}, \rf{C}, we may  express  the
matter content of the space-time in terms  of the  geometric quantity
$b_{ij}$ and the  geometry of $V_{5}$. 

As  a simple example, consider a flat  $V_{5}$.
In this case,  a general expression of  $b_{ij}$ is given by \cite{Szk}:
\[
b_{ij}= \lambda\, g_{ij} +(4\lambda-h)v_{i}v_{j} 
\]
where  $\lambda$ is  an eigenvalue of  $b_{ij}$ with  eigenvector $v_{i}$.
When replaced in  \rf{EE1},  we obtain severe limitations on the type  of
matter that could exist in the four dimensional space-time.

As a  second example  consider a Ricci flat $V_{5}$, $\epsilon=1$,
containing a  4-dimensional space-time filled with dust:
$t_{ij}^{source} = \rho u_{i}u_{j}$,  where  $g_{ij}u^{i}u^{j}=-1$ and  $\rho$
is 
the matter density. In this case \rf{t}  gives the algebraic relation
on $b_{ij}$:
\[
g^{mn}b_{im}b_{jn} -hb_{ij} -\frac{1}{2} (k^{2}-h^{2})g_{ij} - \epsilon\,
\R_{\alpha\beta\gamma\delta}\N^{\alpha}\N^{\delta}\Y^{\beta}_{,i}  
\Y^{\gamma}_{,j}   =  8\pi G\, \rho u_{i}u_{j}
\]
As we see, besides the extrinsic terms of  the geometry of  $V_{4}$,  there is
also a  contribution  from the geometry of  $V_{5}$ to 
the  matter content of the space-time.  

One important case occurs  when
we assume that $V_{5}$  is  subjected to an  additional condition:
\[
\R_{\alpha\beta\gamma\delta}\N^{\alpha}\N^{\delta}\Y^{\beta}_{,i}  
\Y^{\gamma}_{,j}=0.\]  
In this case \rf{EE1} leads to  a  purely algebraic expression 
\[
8 pi G t_{ij}^{\mbox{source}}  = \epsilon\, t_{ij}^{\mbox{geom.}}
\]
Applying to  the  case of dust  we obtain 
\[ k^{2}-h^{2}=\frac{ 8\pi G\, \rho}{\epsilon}\]
which is  an algebraic equation on  $b_{ij}$ with solution
$b_{ij}= \left( h+\frac{8\pi G\,\epsilon\rho}{h}\right)g_{ij}$.
Replacing  the definition  of  $h$    we obtain
\[ 
b_{ij}  = \sqrt{\frac{-2\epsilon \pi G\epsilon\rho}{3}}\,\; g_{ij}
\]
so that  the extrinsic curvature  is  in direct proportion to 
the square root of the  density. 

The  above examples represent very simple situations
which illustrate the consequences  of Campbell's theorem to  general
relativity, namely that  $V_{5}$  has to do with the 
dynamics of space-time rather than being a mere mathematical support.

On the other  hand, the extrinsic curvature $b_{ij}$ is  a  symmetric tensor
in space-time  so that it may  behave as  a  spin 2 field. Since this  enters
as  a  source in  \rf{EE1},   we  do not obtain the
Einstein-Maxwell equations as  we would expect from a five dimensional
Kaluza-Klein theory.  However,  if the number of dimensions of the
embedding space is increased, the twisting vector, emerges  playing
the role of a  gauge field and the  Kaluza-Klein metric   would emerge
naturally from the embedding\cite{Maia:1}.

\end{document}